# Direct detection of induced magnetic moment and efficient spin-to-charge conversion in graphene/ferromagnetic structures

J. B. S. Mendes[1,*], O. Alves Santos[1,2], T. Chagas[3], R. Magalhães-Paniago[3], T. J. A. Mori[4], J. Holanda[2], L. M. Meireles[3], R. G. Lacerda[3], A. Azevedo[2], and S. M. Rezende[2]

[1]*Departamento de Física, Universidade Federal de Viçosa, 36570-900, Viçosa, MG, Brazil*
[2]*Departamento de Física, Universidade Federal de Pernambuco, 50670-901, Recife, PE, Brazil.*
[3]*Departamento de Física, Universidade Federal de Minas Gerais, 31270-901, Belo Horizonte, MG, Brazil.*
[4]*Laboratório Nacional de Luz Síncrotron, Centro Nacional de Pesquisa em Energia e Materiais, 13083-970 Campinas, SP, Brazil*

This article shows that the spin-to-charge current conversion in single-layer graphene (SLG) by means of the inverse Rashba-Edelstein effect (IREE) is made possible with the integration of this remarkable 2D-material with the unique ferrimagnetic insulator yttrium iron garnet (YIG = $Y_3Fe_5O_{12}$) as well as with the ferromagnetic metal permalloy (Py = $Ni_{81}Fe_{19}$). By means of X-ray absorption spectroscopy (XAS) and magnetic circular dichroism (XMCD) techniques, we show that the carbon atoms of the SLG acquires an induced magnetic moment due to the proximity effect with the magnetic layer. The spin currents are generated in the magnetic layer by spin pumping from microwave driven ferromagnetic resonance and are detected by a dc voltage along the graphene layer, at room temperature. The spin-to-charge current conversion, occurring at the graphene layer, is explained by the extrinsic spin-orbit interaction (SOI) induced by the proximity effect with the ferromagnetic layer. The results obtained for the SLG/YIG and SLG/Py systems confirm very similar values for the IREE parameter, which are larger than the values reported in previous studies for SLG. We also report systematic investigations of the electronic and magnetic properties of the SLG/YIG by means of scanning tunneling microscopy (STM).

**KEYWORDS:** graphene, magnetic proximity, magnetic circular dichroism, spin-pumping, spin-to-charge conversion, inverse Rashba-Edelstein effect, extrinsic SOI in graphene.

## I - INTRODUCTION

New phenomena discovered in condensed matter physics and materials science in the last decades have shown that Spintronics has many advantages over conventional electronics and several spin based devices are being used in consumer electronics[1,2]. Nowadays, generation, transport and manipulation of spin currents - i.e., net flows of spin angular momentum - are among the main challenging topics in this research area. In turn, among the main phenomena to generate and manipulate spin currents, one can mention the spin pumping effect (SPE), the spin Hall effect (SHE) and its Onsager reciprocal, the inverse SHE (ISHE)[3–11]. In the SPE, the precessing magnetization in a ferromagnetic material (FM) generates a pure spin current in an adjacent normal metal (NM), and in the SHE, an unpolarized charge current is partially converted into a transverse spin current by the spin-orbit coupling (SOC). Spin current has the advantage of flowing in both metallic and insulating materials that can be magnetic or non-magnetic. This features increase the range of spintronics phenomena that are of interest for basic and applied research, such as: (i) spin transport in magnetic insulators and organics materials[12–17]; (ii) spin-to-charge current-conversion in paramagnetic, ferromagnetic and antiferromagnetic metals, as well as in semiconductors[18–28]; (iii) magnetization manipulation by spin transfer torque (STT)[29–33]; and spin-charge interplay in two dimensional systems based on the Rashba-Edelstein effect[34–40].

From the applications point of view the creation of spin-logic devices, which allow the control and transport of the spin current over long distances, is one of the major research challenges in spintronics. In this regard, graphene - a single atomic layer of carbon atoms in a honeycomb lattice (see Figure 1(c)) - has attracted great attention as a promising material for spin based devices due to its exceptional electronic transport properties, excellent charge carrier mobility, quantum transport, long spin diffusion lengths and spin relaxation times[41–45]. However, due to the low atomic number of carbon, the Dirac electrons in intrinsic graphene have a weak SOC and no magnetic moments, thus, reducing the promises of applications in spintronics[2]. In order to overcome this limitation, several mechanisms and structures have been proposed mainly focused on the improvement of local magnetic moments and spin-orbit coupling in graphene. Among them we may cite: (i) functionalization of graphene with small doses of adatoms or nanoparticles[44,45]; (ii) development of nanostructures with gate electrodes of ferromagnetic materials[46,47]; (iii) and by directly attaching graphene in contact with a ferromagnetic insulator (FMI) or a ferromagnetic metal (FM), to induce magnetic proximity effects[38,48–51]. More recently, it has been shown a large enhancement of the SOC in graphene when it is directly fabricated on top of semiconducting transition-metal dichalcogenides[52–54].

*Corresponding author: Joaquim B. S. Mendes, joaquim.mendes@ufv.br

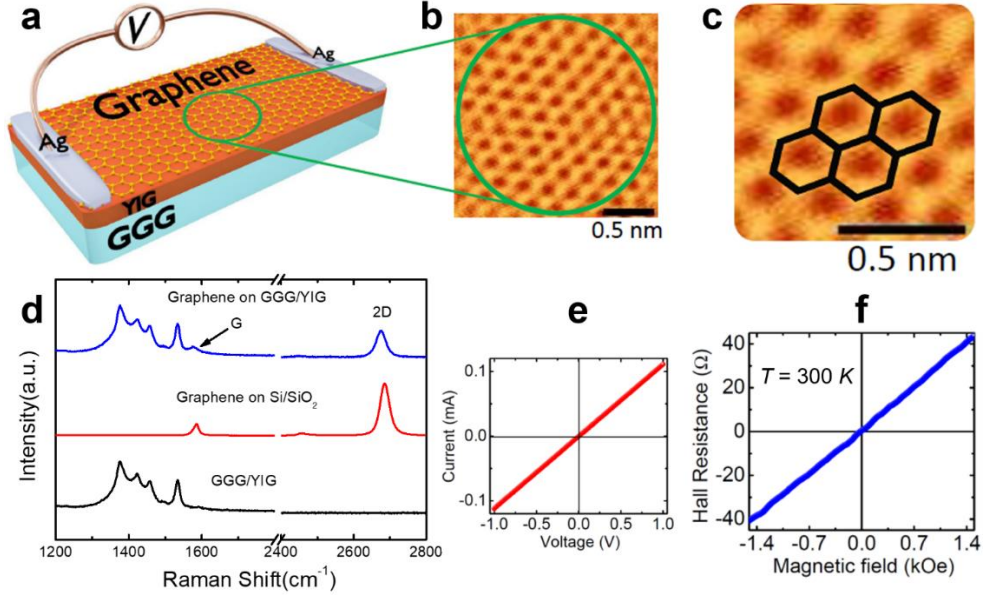

**Figure 1.** (color online) **(a)** Sketch showing the YIG/graphene structures and the electrodes used to measure the dc voltage due to the IREE charge current in the graphene layer resulting from the spin currents generated by microwave FMR spin pumping. **(b)** STM image showing the surface details from an arbitrary region in the SLG/YIG sample. **(c)** High-resolution image showing the typical hexagons of the graphene deposited on YIG. **(d)** Raman spectroscopy spectra of GGG/YIG, SiO$_2$/graphene and GGG/YIG/graphene. **(e)** I-V curve of the YIG/SLG structure demonstrating the formation of Ohmic contacts between the electrodes and the SLG. **(f)** Low-magnetic-field dependence of the Hall resistance at room temperature.

Recent experimental studies have shown that YIG turned out to be an ideal ferromagnetic insulating (FMI) material to induce spin properties in hybrid structures with single-layer graphene (SLG)[38,55–59]. Among other reasons for the use of YIG, we can mention its excellent mechanical and structural properties, as well as its extraordinary low magnetic damping. At the same time, the metallic ferromagnetic Py (Ni$_{81}$Fe$_{19}$) can used instead of YIG if a conducting material is required for device development. In this manuscript, we report an extensive investigation of spin-current to charge-current conversion by inverse Rashba-Edelstein effect (IREE) in large area SLG grown by chemical vapor deposition (CVD) on YIG and Py films. Thus, we show that the SLG in atomic contact with both magnetic materials exhibits an appreciable extrinsic SOC that enhances the IREE. Besides that, as the unequivocal proof of the proximity-induced magnetism in graphene, we performed X-ray magnetic circular dichroism (XMCD) measurements in YIG/SLG and Py/SLG samples. The XMCD results confirmed the existence of an induced magnetic moment at the carbon atoms of graphene.

## II - SAMPLE PREPARATION AND CHARACTERIZATION

In the experiments reported here, we investigated two types of heterostructures. In the first structure, the graphene is interfaced with a ferrimagnetic insulator material (sample A: YIG/SLG), and in the second heterostructure, the graphene is interfaced with a conductive ferromagnetic material (sample B: Py/SLG). As ferrimagnetic insulator material we have used a single-crystal YIG film with thickness 6.0 μm grown by liquid-phase epitaxy on a 0.5 mm thick substrate of (111) gadolinium gallium garnet (GGG). The YIG samples were cut from the same GGG/YIG wafer in the form of rectangles with lateral dimensions 2.5 x 8.0 mm$^2$. The samples were prepared in the following way. The graphene layers were grown on Cu foils inside a CVD chamber at 1000 °C with a mixture of CH$_4$ (33% volume) and H$_2$ (66% volume) at 330 mTorr for 2.5 hours[60,61]. A 120 nm thick layer of PMMA is spun on top of the graphene/Cu surface and then the Cu is etched away. The PMMA/graphene structure is now transferred to the YIG film. Finally, the PMMA is removed using solvents and the sample receives two silver paint contacts as illustrated in Figure 1(a) [See Sup. Mat., Section I and Figure S1, for further details on the growth conditions as well as the transference of graphene]. For sample B preparation, we followed the same procedure, but in this case the graphene is transferred directly to the substrate of SiO$_2$ (300nm)/Si(100). As schematically shown in Figure 5(a), the SLG surface is partially covered by a Py layer leaving the ends free to coat two Ag electrodes. With this configuration, the effect of the Py conductivity does not affect the electrical detection of the spin-current to charge-current that occurs on the SLG. The scanning tunneling microscopy (STM) image of Figure 1(b) shows surface details for the SLG/YIG sample, where it allows us to observe the flat and homogeneous graphene surface. The honeycomb hexagon lattice of the SLG, deposited onto YIG, is clearly seen by the high-resolution STM image shown in Figure 1(c). Additional characterizations of the high-quality YIG layer used in this work can be verified in Fig. S2 of the Sup. Mat., such as X-ray diffraction measurements, surface roughness, magnetic hysteresis, as well as the magnetic domain structures measured by means of the magnetic force microscopy technique.

Figure 1(d) shows the Raman spectroscopy spectra of GGG/YIG, SiO$_2$/SLG and GGG/YIG/SLG. The spectrum at the top of Fig. 1(d), shows the lines of GGG/YIG/SLG. The characteristic peaks from the G (1580 cm$^{-1}$) and 2D (2700 cm$^{-1}$) bands of the SLG confirm the successful transfer of graphene to the surface of the YIG layer. Figure 1(e) shows the measured linear I×V characteristics, indicating Ohmic contacts between the Ag-electrodes and the graphene. The magnetic-field dependence of the Hall resistance can be seen checked in Figure 1(f), in the field range where the magnetization of YIG rotates out of plane.

### III - STM/STS EXPERIMENTS

Scanning tunneling microscopy/spectroscopy (STM/STS) measurements were carried out using a Nanosurf microscope (in constant-current mode) operating at room temperature (300 K) using freshly cleaved Pt-Ir tips. Spectroscopy data were obtained through numerical differentiation of measurements of the tunneling current as a function of the sample bias at selected points (averaged over ten individual STS measurements). The STM images were obtained employing a voltage of 40.0 mV – 70.0 mV and a tunneling current of 1.0 nA – 1.5 nA. In order to investigate the electronic response of the sample under the application of a magnetic field (H), STM/STS measurements were carried out in atomically flat regions as well as in the vicinity of steps and edges. These are regions where the spin orientation of the electrons is changed due to the presence of a magnetic field[62]. A STM image of the sample, where one observes a region of low roughness with the presence of some steps, is shown in Fig. 2(a). STS spectra, Fig. 2(b) were measured at the points marked with the green circles in the STM image and labeled as 1 (left plot), i.e., at a flat region and 2 (right plot), i.e., at a step edge. Both measurements were repeated with and without the presence of a magnetic field parallel to the sample surface. In this figure the black and the red curves, in both panels, are related to STS spectra measured for H = 0.0 Oe and H = 10.0 Oe, respectively. In both conditions a V-shaped spectrum, compatible with the typical electronic signature of a monolayer graphene[43], is observed. In this measurement one also observes a natural broadening of $4K_BT \sim 100$ meV, near the Dirac point, which is a consequence of temperature of the experiment (300 K)[63]. This result evidences that the graphene layer is decoupled from the substrate for this particular region since its local density of states (LDOS) is not significantly affected by the substrate.

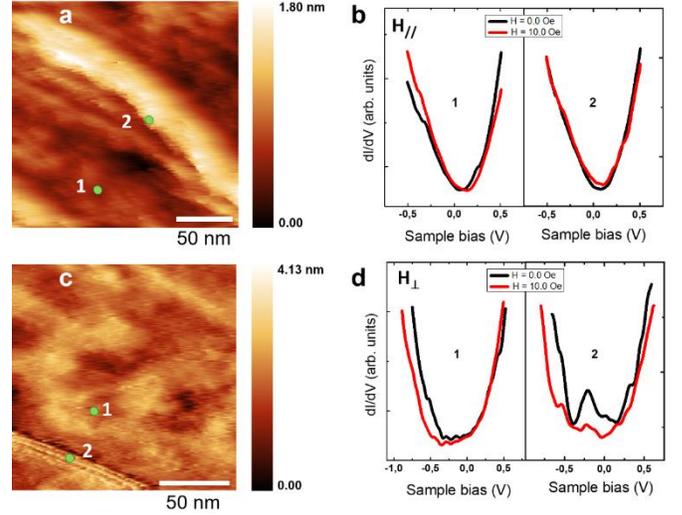

**Figure 2.** (color online) **(a)** STM topography image of the graphene/YIG sample showing a region of low roughness. **(b)** STS spectra acquired at the points marked with the green circles (shown in a) and labeled as 1 (left panel) and 2 (right panel). The black curves were measured in the absence of a magnetic field and the red curves were measured for a magnetic field of 1.00 mT parallel to the sample surface. **(c)** STM image showing an atomically flat region with some steps in the left lower corner. **(d)** STS spectra acquired at the points marked with the green circles (shown in c) and labeled as 1 (left panel) and 2 (right panel). The black curves were measured in the absence of a magnetic field and the red curves were measured for a magnetic field of 1.00 mT perpendicular to the sample surface. Scale bars are: (a) 50 nm and (c) 40 nm.

On the contrary, for Figs.2(c) and 2(d) a different scenario is observed. In Fig.2(c) an atomically flat region with the presence of some steps is presented. Differently from the previous case, for this area of the system a magnetic field perpendicular to the sample surface was applied. STS spectra for points 1 (flat region) and 2 (step region) indicated in this figure are shown in Fig. 2(d), in the left and right panels, respectively. For the atomically flat region marked as 1, the STS spectra slightly differ from the ones shown in the left panel of Fig. 2(b). In this last case the STS spectra observed do not present a completely linear relationship between LDOS and energy, which indicates a higher degree of coupling of the graphene sheet to the substrate. In addition, no significant changes are observed due to the application of H. For the STS spectra acquired at point 2 (shown in the right panel of Fig. 2(d)) a distinct electronic behavior is noticed. Analyzing the LDOS measured at the step without the application of a magnetic field one observes a peak at -0.25 V, which is probably related to an edge state. Interestingly, this peak is highly suppressed in the presence of a magnetic field of 10.0 Oe showing that the LDOS is influenced by H under this condition. This effect may be directly related to the change of the magnetoresistance for different values of magnetic fields reported in reference [38]. Thus, this result indicates that the changes observed in the magnetoresistance for H = 0.0 Oe and H = 10.0 Oe must be are related to the local electronic density of edge states.

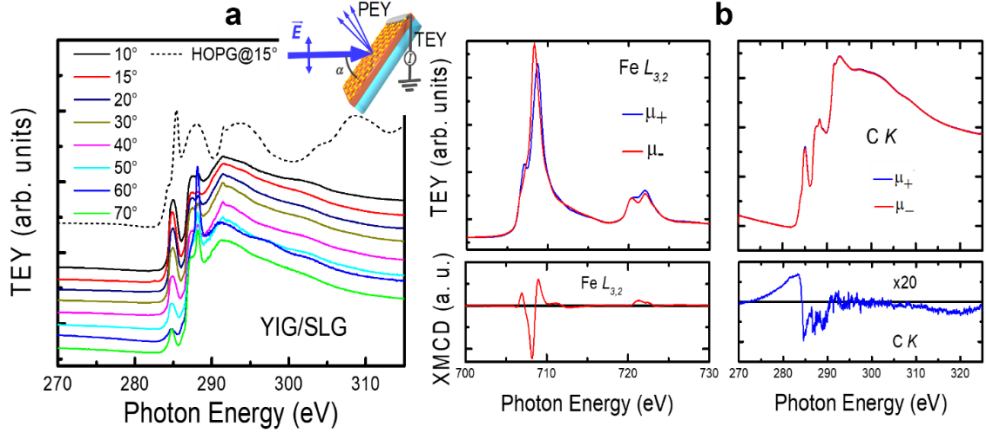

**Figure 3.** (color online) **(a)** Spectra acquired around the carbon K edge of the YIG/graphene, with the electric field vector of the linearly polarized beam pointing along the vertical direction, for several incidence angles α. The dash-lined spectrum is acquired from a HOPG reference sample with an incidence angle of 15°. **(b)** XAS spectra acquired around the Fe L3,2 (left) and C K (right) edges, for both right and left-handed circular polarizations of the light, under an applied magnetic field of 300 Oe and at an incidence angle of 15°. The bottom of the figures presents the XMCD asymmetries.

## IV - XAS AND XMCD EXPERIMENTS

Electronic orbital orientation and element-specific magnetic characterization were performed by X-ray absorption spectroscopy, i.e., X-ray linear (XLD) and magnetic circular dichroism, respectively. The measurements were carried out at the U11A-PGM beamline of the Brazilian Synchrotron Light Laboratory (LNLS) near both the carbon K and the iron $L_{3,2}$ absorption edges, at room temperature by total electron yield (TEY) mode. In order to determine the average spatial orientation of the molecular orbitals (π or σ) at the YIG/SLG interface, several XAS spectra were acquired by keeping the electric field of the linearly polarized light pointing along the vertical direction whilst the sample was tilted so that the angle between the electric field and the surface normal was varied from 10° to 90° (see inset of Figure 3 (a) for details on the geometry). To check the magnetism at both carbon and iron absorption edges, the XAS spectra were collected under four configurations of applied magnetic field and polarization of light (+/-300 Oe and right/left-handed circular polarization), with a grazing incidence angle of 15°. Although it is faster to change polarization than magnetic field at the end station/beamline utilized for the experiment, the strong X-ray natural circular dichroism (XNCD) due to the chirality of graphene requires the inversion of the magnetic field to get rid of the contribution of the XNCD to the circular dichroism. Therefore, in order to get the best quality data, the XMCD asymmetry spectra were obtained by taking the average of the differences (modulus) between the spectra acquired for two different polarizations for each magnetic field direction (XMCD = [|TEY_($H_+$, $\mu_+$) − TEY_($H_+$, $\mu_-$)| + TEY$_{H_-,\mu_+}$ − TEY_($H_-$, $\mu_-$)|] / 2 where $H_+$/$H_-$ and $\mu_+$/$\mu_-$ stand for the direction of the magnetic field and for the hand of the circular polarization, respectively).

The average spatial orientation of the molecular orbitals at the YIG/SLG interface can be addressed by acquiring several X-ray absorption spectra for different orientations of the sample out-of-plane direction with respect to the electric field direction of the linearly polarized X-ray beam. In this experiment, the absorption intensity of a specific orbital final state presents a maximum if the electric field is parallel to the direction of maximum density of states of the molecular orbital. On the other hand, the absorption intensity vanishes if the electric field is perpendicular to the molecular orbital axis. Figure 3(a) presents spectra acquired near the carbon K edge, with the electric field vector of the linearly polarized beam pointing along the vertical direction, for several incidence angles $\alpha$. In the geometry illustrated in the inset of the figure, the electric field of the light is along the surface normal at $\alpha = 0°$, and parallel to the surface at $\alpha = 90°$. The changes of the XAS lineshape show that the spectral feature around 285 eV, which is assigned to the C 1s→π* transition of 1s core electrons into unoccupied π* states, is remarkably strong for grazing incidence angles and tends to vanish as the incidence angle is increased towards normal incidence. This strong angular dependence of the XAS lineshape is due to different orbital orientations and demonstrates that the π* unoccupied states are aligned out-of-plane, as it is expected for the C $p_z$ orbitals in the graphene-type layered $sp^2$ coordination. This agrees with the structural characterization by STM and also proves the good quality of the SLG.

It is worth noting the spectral features around 286-290 eV, which is ascribed to the PMMA presence on the sample surface (residue of the graphene transferring process) and jeopardizes the analysis of the spectra as it is in the very middle region between the spectral features due to C 1s→π* (283-289 eV) and C 1s→ σ* (289-315 eV) transitions. A comparison with a spectrum around the C K edge of highly oriented pyrolytic graphite (dash-lined spectrum in Figure 3(a), acquired from a HOPG sample with an incidence angle of 15°) allows one to notice changes in the lineshape of the YIG/SLG spectrum. These changes were attributed to chemisorption as suggested in references[64] and[65]. Small shifts in energy and remarkable broadening of the π* and σ* resonances suggest orbital hybridization and electron sharing along the interface, with delocalization of the corresponding core-excited state. Such a covalent interfacial bonding between carbon atoms in the SLG and iron atoms in the YIG could be the origin of the magnetic proximity effect detected in carbon atoms. In a very similar way as it has been demonstrated for the system Ni(111)/graphene[64,65], we suggest that a hybridization between graphene π and valence-band states at the YIG/graphene interface leads to partial charge transfer of spin-polarized electrons from the substrate to C atoms in graphene. In this sense, an induced magnetic moment of carbon atoms can be detected by XMCD, as shown in Figure 3(b).

Figure 3(b) shows XAS spectra acquired around the Fe $L_{3,2}$ (left) and C K (right) edges, for both right and left hand circular polarizations of light, under an applied magnetic field of 300 Oe and at an incidence angle of 15°. The bottom of the figures present the XMCD asymmetries, which were evaluated by taking the averaged difference between the difference of the spectra taken with two different polarizations under 300 Oe (shown in the figure) and the difference of the spectra taken under -300 Oe (not shown), as explained in the methods section. The Fe $L_{3,2}$ XAS/XMCD spectrum is in agreement with published spectroscopic data of iron(III) coordinated as it is in YIG[66]. The most important result of the XAS experiments is the clear observation of a dichroic signal at the C K absorption edge. Despite the presence of the PMMA feature in the XAS spectra, it is clear in the XMCD spectrum that the asymmetry has its maximum at the 1s→π* resonance. In other words, the major magnetic response of the carbon atoms arises from transitions of the 1s electrons into the unoccupied π* states, the transitions to the σ* states yielding a very weak magnetic signal. This result, also in accordance with references[64] and[65] for Ni(111)/graphene, suggests that mostly the C 2$p_z$ orbitals of the SLG are magnetically polarized due to the hybridization with the valence-band states of the YIG substrate. In reference[67] the authors report on magnetic signals of carbon observed by the X-ray resonant magnetic reflectivity technique, the ferromagnetism of carbon in the Fe/C multilayered system being related to the hybridization between the Fe 3d band and the C $p_z$ orbitals. Besides, there are other reports on magnetism in the Fe/C interface investigated by XMCD and on the magnetism induced in carbon nanotubes on ferromagnetic Co substrate due to spin-polarized charge transfer at the interface[68,69].

The XAS and XMCD spectra of a sample Py/SLG were also acquired (see Figure S3 in the supplementary materials). Similar to the results obtained for the sample YIG/SLG, the XAS spectra for the Py/SLG, measured with linear polarization and varying the incidence angle, confirm the good quality of that sample, with the π* unoccupied states of carbon pointing out-of-plane. In its turn, the circular dichroism measured at carbon K edge also present a remarkable asymmetry around the feature related to the π* unoccupied states. Combined with the XMCD spectra at the Ni $L_{3,2}$ the results also suggest that the C $2p_z$ orbitals of the SLG are magnetically polarized due to the hybridization with the valence-band states of the adjacent Py layer.

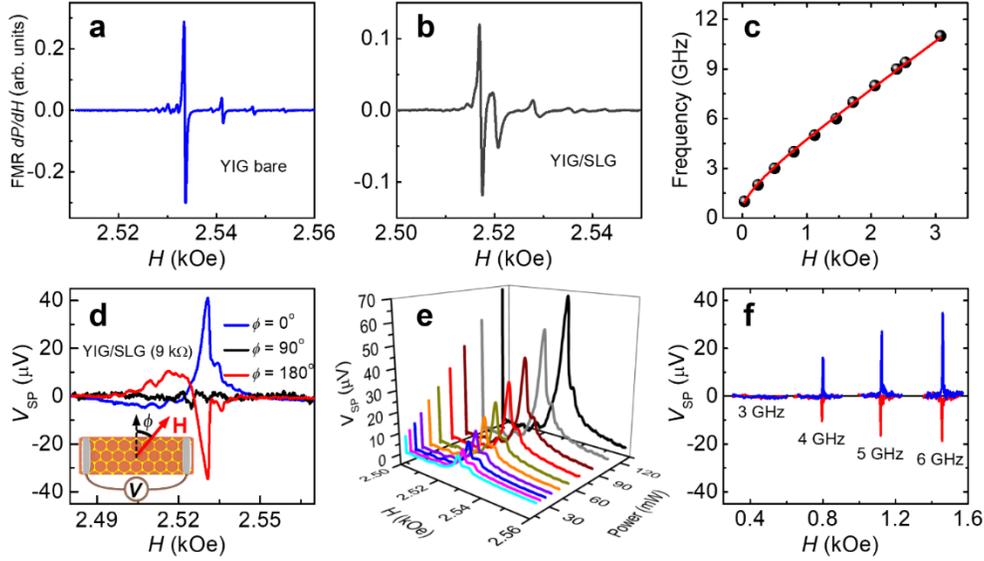

**Figure 4.** (color online) **(a)** and **(b)** shows the field scan FMR microwave absorption derivative spectra for YIG and YIG/SLG respectively, with the magnetic field is applied in the film plane, normal to the long dimension. **(c)** Driving frequency versus FMR resonance field, where the solid red line corresponds to best fit of Kittel equation. **(d)** Field scan of dc spin pumping voltage measured with 9.4 GHz microwave driving with power of 80 mW in YIG/SLG for three different directions between in plane magnetic field and detection electrodes. **(e)** Field scan dc SPE voltage for different input microwave power. The results shows the linear variation of the peak voltage with microwave power in YIG/SLG. **(f)** SPE voltage for four different driving microwave frequencies for $\phi = 0°, 180°$.

## V - SPIN-PUMPING EXPERIMENTS

In the spin pumping experiments, the FM /SLG bilayer sample is under action of both a small radio-frequency (rf) magnetic field applied perpendicular to a dc magnetic field in the ferromagnetic resonance (FMR) configuration. The precessing magnetization $\vec{M}$ in the FM layer generates a spin current density at the FM/SLG interface, $\vec{J}_S = \left(\hbar g_{eff}^{\uparrow\downarrow}/4\pi M^2\right)\left(\vec{M} \times \frac{\partial \vec{M}}{\partial t}\right)$, where $g_{eff}^{\uparrow\downarrow}$ is the real part of the effective spin mixing conductance at the interface that takes into account the spin-pumped and back-flow spin currents[6]. The net spin current that flows into the adjacent conducting layer produces two effects: (i) increases the damping of the magnetization due to the flow of spin angular momentum out to the adjacent material; (ii) due to the inverse Rashba Edelstein effect (IREE), the spin current that flows out of the FM layer is converted into a perpendicular charge current that produces a dc-voltage at the ends of the graphene layer, measured between the two Ag electrodes. By investigating, these two independent phenomena it is possible to extract material parameters from the measurements of the FMR absorption and the spin-pumping voltage.

In the microwave FMR absorption and spin-charge conversion experiments the sample with electrodes, show in Fig. 1(a) and 5(a), is mounted on the tip of a PVC rod and inserted into a small hole in the back wall of a rectangular microwave cavity operating in the $TE_{102}$ mode, in a nodal position of maximum rf magnetic field and zero electric field. For sample A, we did not use a microwave cavity because the detuning of the resonance at the strong FMR absorption of YIG introduces distortions in the lineshapes. The waveguide is placed between the poles of an electromagnet so that the sample can be rotated while maintaining the static and rf fields in the sample plane and perpendicular to each other. With this configuration we can investigate the angular dependence of the FMR absorption spectra. Field scan spectra of the derivative $dP/dH$ of the microwave absorption are obtained by modulating the dc field with a weak ac field at 1.2 kHz that is used as the reference for lock-in detection. Figures 4(b) and 5(b) show the derivative of the FMR spectra of the two samples investigated in this work. Both spectra were obtained with $H$ in-plane and normal to the long sample dimension, frequency $f = 9.4$ GHz and input microwave power of 80 for sample A and 24 mW for sample B. In Fig. 4(a), for a bare YIG film, one clearly sees the various lines corresponding to standing spin-wave modes with quantized wave numbers $k$ due to the boundary conditions at the edges of the film. The strongest line corresponds to the uniform (FMR) mode that has frequency close to the spin-wave mode with $k = 0$, the lines to the left of the FMR correspond to hybridized standing spin-wave surface modes whereas those to the right are volume modes. All modes have similar peak-to-peak linewidth of 0.39 Oe, corresponding to a half-width-at-half-maximum (HWHM) of $\Delta H \approx 0.34$ Oe of direct microwave absorption. Fig 4(b) show the FMR absorption for YIG/SLG sample. Due to the spin injection into graphene the linewidth of YIG/SLG increases up to $\Delta H_{YIG/LSG} = 0.90$ Oe. As can be seen in Fig. 4(b), the contact of the YIG film with the graphene layer increases the FMR linewidth of YIG. From the line broadening we can obtain a good estimate for the spin mixing conductance $g_{eff}^{\uparrow\downarrow}$, using the relation $g_{eff}^{\uparrow\downarrow} = \frac{4\pi M_0 t_{FM}}{\hbar \omega}(\Delta H_{FM} - \Delta H_{FM/SLG})$, from which we obtained $g_{eff}^{\uparrow\downarrow} \approx 10^{14}$ cm$^{-2}$ from the YIG/SLG interface.

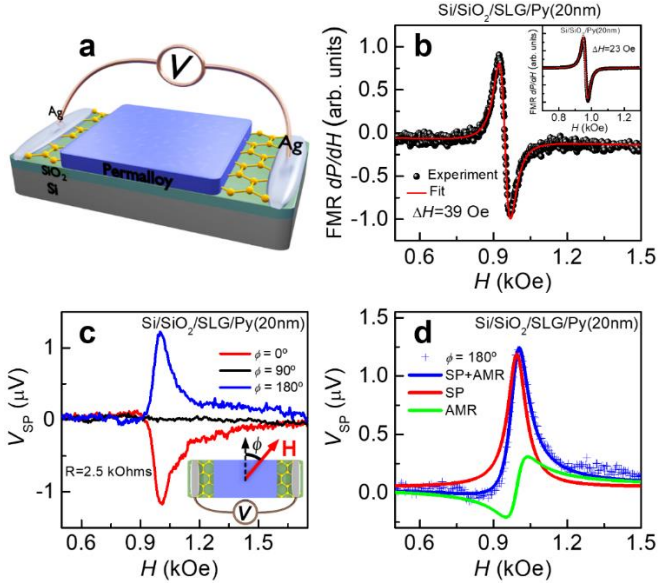

**Figure 5.** (color online) **(a)** Sketch showing the Si/SiO2/SLG/Py(20 nm) structure. **(b)** Field scan of derivative power absorption of Si/SiO2/SLG/Py(20 nm). The best fit was adjusted with linewidth of 39 Oe, and the inset show FMR absorption of Py on top of Si with $\Delta H = 23$ Oe. **(c)** dc spin pumping voltage in SLG/Py. **(d)** The blue line represents a sum of a Lorentzian (symmetric red line) and a Lorentzian derivate (antisymmetric green line) best fit form data of $V_{SP}$ ($\phi = 180°$).

Figure 4(c) shows the FMR measurements carried at several microwave frequencies employing a FMR broadband spectrometer. The microwave excitation was carried out by means of a microstrip transmission line of copper, 0.5 mm wide and characteristic impedance $Z_0 =$ 50 W that was fabricated on a Duroid substrate, where the return is the ground plane on the back side of the board. The YIG film placed on top of the microstrip line separated by a 60 μm thick Mylar sheet is excited by the rf magnetic field perpendicular to the static field $H$ applied in the film plane. By keeping the frequency value fixed and sweeping the static magnetic field, the variation of the transmitted power due to the FMR absorption is detected by a Schottky diode. The applied magnetic field is modulated with frequency 8.7 kHz and amplitude 0.45 Oe so as to allow lock-in detection of the field derivative of the power absorption. The solid red line in the Fig.4 (c) was obtained from the fit with Kittel equation, $f = \gamma[(H_R - H_A)(H_R + H_A + 4\pi M_S)]^{1/2}$, where $\gamma = g\mu_B/\hbar$ ($\approx 2.8\ GHz/kOe$) is the gyromagnetic ratio, $g$ is the spectroscopic splitting factor, $\mu_B$ the Bohr magneton, $\hbar$ the reduced Plank constant, $H_A$ the in-plane anisotropy field, and $4\pi M_S = 1760$ G the saturation magnetization. With the fit shown in Fig. 4(c) we obtained $H_A = 28\ Oe$.

The net spin current that reaches the YIG/SLG interface creates a spin accumulation on SLG, this accumulation can be converted into charge current, we can write the dc spin pumping spin-current density at YIG/SLG interface generated by magnetization precession as

$$J_S(0) = \frac{\hbar \omega p g_{eff}^{\uparrow\downarrow}}{16\pi}\left(\frac{h_{rf}}{\Delta H}\right)^2 L(H - H_R), \quad (1)$$

were $\Delta H$ and $H_R$ are the linewidth and field for resonance of YIG/SLG bilayer, $L(H - H_R)$ represents the Lorentzian function, $p$ is the precession ellipticity and $\omega = 2\pi f$ and $h_{rf}$ are the frequency and amplitude of the driving microwave magnetic field, respectively[18]. This spin current is converted into a transversal charge current, at the SLG by means of the IREE, which can be detected by measuring a dc voltage that sets up between the two Ag electrodes. The dc voltage is measured by a nanovoltmeter directly connected to the edges of sample as sketched in Figures 1(a) and 5(a). The direct measurement of the spin-pumping voltage $V_{SP}(H)$ spectra are obtained by sweeping the dc magnetic field with no AC field modulation. By rotating the sample in the plane, we measured the angular dependence of the voltage as a function of the angle $\phi$ (see inset Fig. 4(d)). Figure 4(d) shows the spectra of $V_{SP}(H)$ obtained with microwave frequency $f = 9.4$ GHz and incident power of 80 mW, for three field directions, $\phi = 0°, 90°, 180°$. The spectra exhibit a large peak at the FMR field position and lateral small peaks corresponding to the standing spin-wave modes. Moreover, this intensity is proportional to pumped spin current, the signal reverses at 180° and fall to noise level in 90°. It is important to mention that the voltage measured in Fig 4(d) with micro-wave power of 80 mW is almost one order of magnitude greater than the values reported in Ref [38], where it was applied a rf power of 150 mW. The asymmetry between the positive and negative peaks is similar to that observed in other bilayer systems and can be attributed to a thermoelectric effect[70,71]. Fig 4(e) shows the field scan of spin pumping voltage spectra for different input microwave power obtained for the $\phi = 0°$ configuration. Note that the amplitude of the peak voltage in YIG/SLG has a linear dependence with the microwave power, showing that nonlinear effects are not being excited in the power range of the experiments.

## VI - SPIN-TO-CHARGE CURRENT CONVERSION RESULTS

Figure 6(a) shows the schematic illustration of a spin current generation driven by the spin pumping effect, where the charge current $\vec{J}_C$ generated at the SLG is simultaneously orthogonal to the dc magnetic field $\vec{H}$ and to the spin current $\vec{J}_S$, for either YIG or Py. In systems as YIG/SLG[38], Ag/Bi[34] and LAO/STO[35], the presence of two-dimensional SOC driven by the Rashba effect[72], as illustrated in Fig 6(b). This effect creates a spin-split in the energy dispersion surfaces of the Rashba states, thus, causing a locking between the momentum $k$ and the spin angular of the electron. In this case, a 3D spin current overpopulates states on one side of the Fermi contour and depopulates states on the other, thereby, creating a shift $\Delta k$ on the Fermi contours which is equivalent to a charge current. This effect is called inverse Rashba-Edelstein effect[34], as illustrated in Figure 6(c), where a spin current ($-z$ direction) with spin polarization ($-x$ direction) creates a charge current ($-y$ direction). Using the relation[34] $j_C = (2e/\hbar)\lambda_{IREE}J_S$, between the 3D spin current in Eq. (1) and 2D charge current in SLG we can estimate the $\lambda_{IREE}$ which is a coefficient characterizing the IREE, with dimension of length and proportional to the Rashba coefficient, and consequently to the magnitude of the SOC in graphene. The charge current density produces a measured voltage described as $V_{IREE} = R_S w j_C$. Using Eq. (1) one can express the IREE coefficient in terms of the measured voltage peak value by,

$$\lambda_{IEE} = \frac{4V_{IREE}^{peak}}{R_S w e f p g_{eff}^{\uparrow\downarrow}(h_{rf}/\Delta H)^2}, \quad (2)$$

where $R_S$ is the shunt resistance and $w$ is the width of SLG. Since YIG is insulating, here $R_S$ is the resistance of the graphene layer. Computing the equation (2) with the parameters: $V_{IREE}^{peak} = 40\ \mu V$, $R_S = 9$ k$\Omega$, $w = 3$ mm, $h_{rf} = 3.5 \times 10^{-2}$Oe and $p = 0.3$, we obtain $\lambda_{IREE} \approx$ 0.002 nm, which is about twice the value measured for the graphene reported in a previous paper[38]. On the other hand, this value is about 2 orders of magnitude smaller than the one obtained for the Ag/Bi interface[34], and about one order of magnitude smaller than the measured for the topological insulators[39,40]. Figure 4(f) shows the measurement of the spin-pumping voltage with scanning $H$ for several frequencies. Here amplitude of the spin-pumping voltage falls abruptly as the frequency decreases below 4 GHz. The reason is that in the YIG films with thickness in the μm range[73] the three-magnon nonlinear process have the coincidence of frequencies at power levels high, as 100 mW[74].

In a further investigation, we replaced the injector of spin current by a metallic ferromagnetic material. In this case, a layer of Py with 20 nm of thickness was deposited on top of a SLG transferred on Si/SiO(300 nm) substrate. Fig 5(a) shows the sketch of the heterostructure

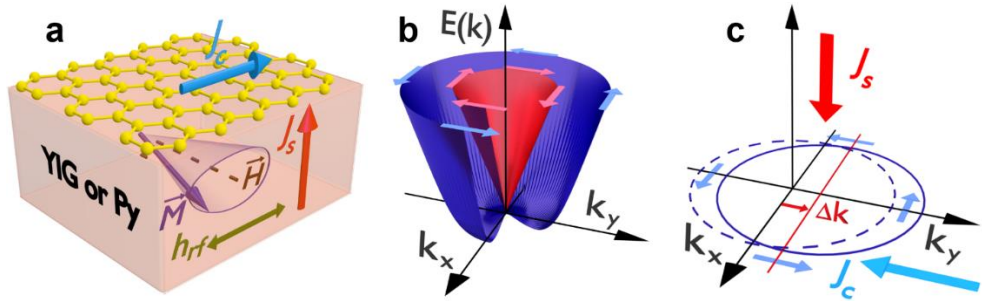

**Figure 6.** (color online) **(a)** Schematic of the Spin-to-charge conversion by the inverse Edelstein effect in FMR spin-pumping experiments on the graphene. **(b)** The spin-polarized band structure (electron energy E as a function of in-plane momentum k) of 2D electron states at a Rashba interface. **(c)** The unbalanced spin states resulting from the 3D spin current creates a 2D charge current in the contour of Rashba interface, i.e inverse Edelstein effect.

Si/SiO/SLG/Py with lateral dimensions of 3 × 1.5 mm. Once again, the sample was mounted on the tip of a PVC rod and inserted through a hole in the center of a back wall of a rectangular microwave cavity operating in the $TE_{102}$ mode, at a frequency of 9.4 GHz with a Q factor of 2000. Fig. 5(b) shows the FMR absorption, at 24 mW, for SiO$_2$/SLG/Py(20 nm). The FMR lineshape was fit with the derivative of a Lorentzian function and the extracted FMR linewidth was 39 Oe. The inset of Fig. 5(b) shows the FMR absorption spectrum for an identical Py layer deposited on Si substrate, with a linewidth of $\Delta H_{Py} = 23$ Oe. As occurred in YIG/SLG, the broadening of the FMR linewidth of Py in contact with graphene layer produces an additional damping due the spin pumping. We obtained $g_{eff}^{\uparrow\downarrow} \approx 10^{15}$cm$^{-2}$ for the SLG/Py interface, which is a value similar to the one observed in Py/Pt interface[19,20,75]. The spin pumping voltage shown if Fig 5(c) was measured by exciting the FMR with the same microwave power. The scan field spectra exhibit the same feature as the $V_{SP}$ measured in YIG/SLG, so that the voltage peak occurs at the FMR field value. However, the direction of spin injection into SLG/Py it's opposite in comparison of YIG/SLG, once that the ferromagnetic material is now on top of graphene, therefore was measured an opposite spin pumping voltage for the same direction, i.e., the maximum positive value occurs for $\phi = 180°$, negative for $\phi = 0°$, and null value for $\phi = 90°$. The field dependence voltage $V(H)$ in the case of Py, can be adjusted by the sum of two components, as shown in Fig. 5(d), where the solid blue line of Fig 5(c), was adjusted by $V(H) = V_{sym}L(H - H_R) + V_{asym}D(H - H_R)$. Here $L(H - H_R)$ is the (symmetric - solid red line) Lorentzian function and $D(H - H_R)$ is the (antisymmetric – solid green line) Lorentzian derivative centered about the FMR resonance field. Taking into account, the amplitude of the microwave field in Eq (2) in Oe is associated to the incident power, in W, by $h_{rf} = 1.776(P_i)^{1/2}$, $R_S = 2.5$kΩ, $w = 1.5$mm, and $V_{sym} = V_{IEE}^{peak} = 1.14$ $\mu$V which corresponds to the amplitude of symmetric component of $V(H)$, we obtain $\lambda_{IREE} \approx 0.003$ nm, very similar with the previous value for YIG/SLG. The antisymmetric component $V_{asym} = 0.52$ $\mu$V is related to the contribution of the anisotropic magnetoresistance (AMR) that occurs on the Py and reflects on the edges of graphene layer[18,76].

## VII - CONCLUSIONS

In this investigation we have shown that the carbon atoms of single layers of graphene acquires an induced magnetic moment due to the proximity effect with ferromagnetic surfaces. This astonishing result was confirmed by means of X-ray absorption spectroscopy (XAS) and magnetic circular dichroism (XMCD) techniques, which shows that the C 2p$_z$ orbitals of the SLG are magnetically polarized due to the hybridization with the valence-band states of the YIG and Py substrates. We have also demonstrated the optimal conversion of a spin current into a charge current in large area of single layers of graphene deposited on YIG and Py films, which is attributed to the inverse Rashba-Edelstein effect (IREE). The results obtained for the SLG/YIG and SLG/Py systems revealed very similar values for the IREE parameter, which are larger than the reported values in the previous studies for SLG[38]. We believe that the spin-to-charge conversion efficiency can be attributed to the excellent structural and interface quality of SLG/FM heterostructures produced, which optimizes the transfer of spins. With this investigation we are shedding light to the very amazing phenomenon of the induced magnetization in the carbon atoms of graphene in contact with magnetic materials and definitely enforces the entrance of graphene as a promising material for basic and applied investigation of spintronics phenomena.

## ACKNOWLEDGMENTS

The authors thank Laboratório Nacional de Luz Síncrotron (LNLS) for providing the beam time (Proposal 20160219, PGM beamline). This research was supported by Conselho Nacional de Desenvolvimento Científico e Tecnológico (CNPq), Coordenação de Aperfeiçoamento de Pessoal de Nível Superior (CAPES), Financiadora de Estudos e Projetos (FINEP), Fundação de Amparo à Pesquisa do Estado de Minas Gerais (FAPEMIG), and Fundação de Amparo à Ciência e Tecnologia do Estado de Pernambuco (FACEPE).

# SUPPLEMENTAL MATERIAL

## I. Conditions for growth and transferring of graphene

The sample A (YIG/SLG) was prepared in the following way. Graphene is grown on Cu foils inside a CVD chamber at 1000 °C with a mixture of $CH_4$ (33% volume) and $H_2$ (66% volume) at 330mTorr for 2.5 hours. A 120 nm thick layer of PMMA is spun on top of the graphene/Cu surface and Cu is then etched away[60,61]. The PMMA/graphene structure is then transferred onto the YIG film, with lateral dimensions of 2.5 x 8.0 $mm^2$. Finally, the PMMA is removed and the sample receives two silver paint contacts as illustrated in Fig. 1(a). Figure S1 shows all the steps involving the transfer of graphene. In the case of Sample B ($Si/SiO_2/SLG/Py$) the graphene is transferred onto $Si/SiO_2$ substrate with lateral dimensions of 1.5 x 3.0 $mm^2$, by the same steps of the previous systematics.

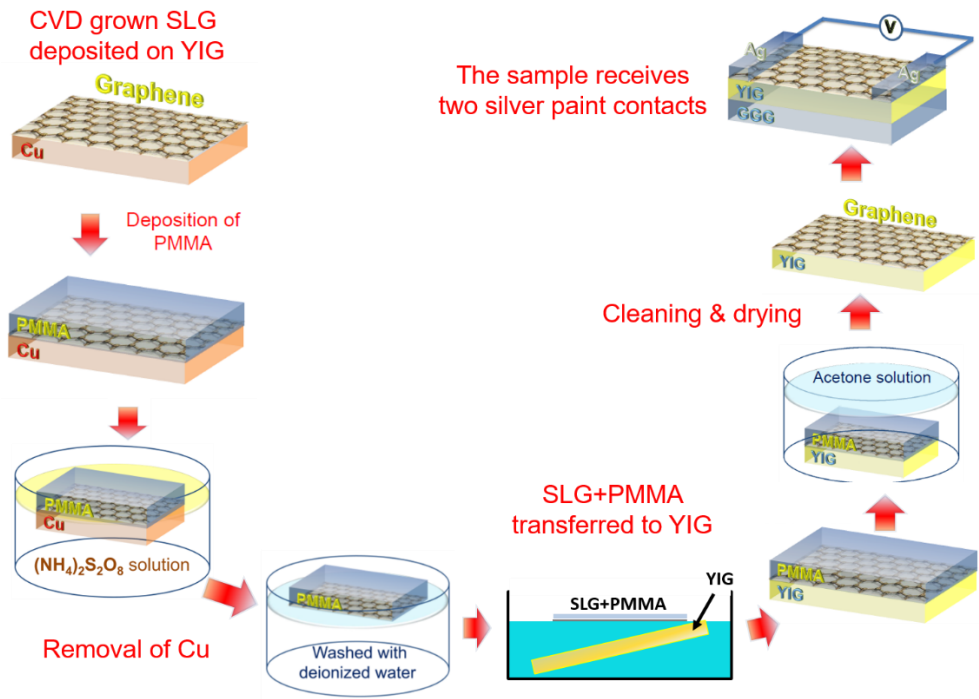

**Figure S1**. (color online) The sequence of steps made in order to insure the best transference of the single-layer-graphene to YIG. Using a similar process, the SLG is transferred to $Si/SiO_2$.

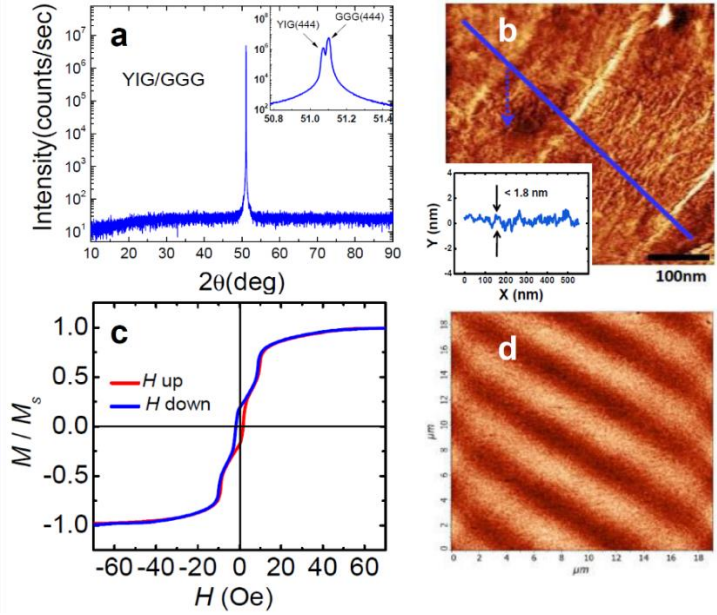

**Figure S2**. (color online) (a) Out-of-plane XRD patterns (ω-2θ scans) of YIG film grown on GGG substrate. The XRD spectrum detailing the position of the peaks of the YIG film and the GGG substrate is shown in the insert. (b) AFM image of the YIG film surface. The line trace in (b) confirms that the YIG surface has a very small roughness. Figure (c) shows a VSM hysteresis curve for the YIG samples used in this work. (d) MFM image of the as-grown 6-$\mu$m-thick YIG film showing up- and down-oriented domains as regions of light and dark contrast.

After this process is deposited a rectangle of Py film with a thickness of 20 nm at mid part of graphene, using a 1.5 x 2.0 mm² shadow mask, as seen in the Fig 5(a). Once again two silver paint contacts are made on the edges of graphene.

## II. Additional information about the sample preparation

The YIG samples consist in a single-crystal yttrium iron garnet (111) films, with thickness of in a 6 $\mu$m, grown by liquid phase epitaxy (LPE) onto a 0.5mm thick (111)-oriented $Gd_3Ga_5O_{12}$ (GGG) substrates. It is important to emphasize that the YIG films were grown by standard LPE technique from a supersaturated melt in which $Y_2O_3$ and $Fe_3O_3$ are added to a flux of $PbO+B_2O_3$. The quality of the YIG samples is attested by the very small FMR linewidth, which is 0.3 Oe (See Fig. 4(a)). Any composition different from $Y_3Fe_5O_{12}$ or phase different from the garnet phase would be impossible to provide such nice small linewidth. In turn, the Py films were deposited by DC sputtering on $SiO_2$/SLG with base pressure of $2.0 \times 10^{-7}$ Torr, and the deposition pressure corresponds to 3 mTorr argon atmosphere in the sputter-up configuration, with the substrate at a distance of 9 cm from the target. A 10 min pre-sputtering procedure was used for cleaning of the targets.

## III. The X-ray diffraction, magnetic hysteresis, and AFM/MFM characterizations in the YIG films

The crystallographic structure of the YIG was assessed by X-ray diffraction (XRD) measurements. The out-of plane XRD pattern is shown in Fig. S2(a). The diffraction patterns were obtained at angles between 10° and 90° (2θ). Only the line associated with the (444) crystal plane of YIG appears in the out-of-plane XRD pattern, indicating that no impurity phases precipitated. The XRD spectrum at high resolution detailing the position of the peaks of the YIG film and the GGG substrate is shown in the insert. The results of XRD measurements indicate that the present YIG film is epitaxially grown on the GGG substrate. The XRD patterns were recorded using the Bruker D8 Discover Diffractometer equipped with the Cu Kα radiation (λ=1.5418 Å).

Figure S2(b) shows an atomic force microscopy (AFM) image of the YIG film surface. The line trace in Fig. S2(b) confirms the uniformity of the YIG film surface with very small roughness. Figure S2(c) shows a magnetization hysteresis curve for a representative YIG sample measured by VSM (Vibrating Sample Magnetometer). The data show a small hysteresis, with a coercive field 1.8 Oe and remanent magnetization 0.12 of the saturation value. Figure S2(d) shows an magnetic force microscopy (MFM) image for the demagnetized state in the 6-$\mu$m thick YIG film, which exhibits a stripes domain structure typical. All the measurements were obtained at room temperature.

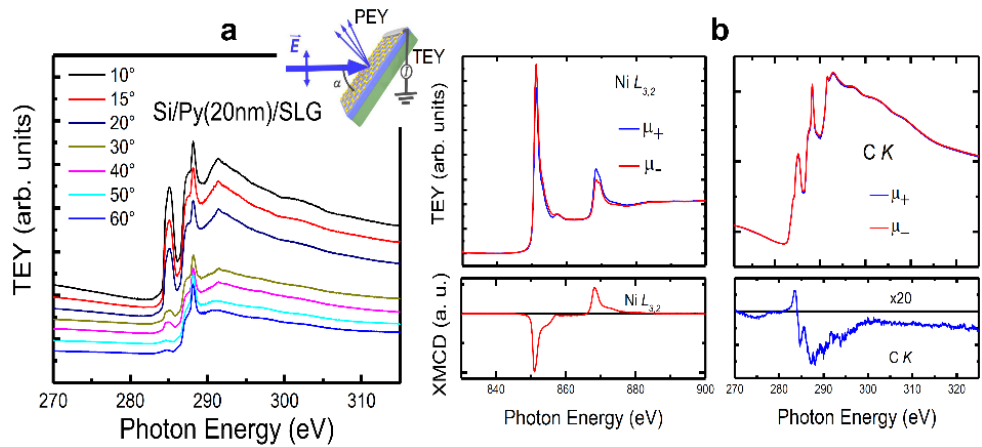

**Figure S3**. (color online) (**a**) XAS spectra of Py/graphene measured at α=15°. (**b**) XAS spectra acquired around the Ni $L_{3,2}$ (left) and C K (right) edges, for both right and left-handed circular polarizations of the light, under an applied magnetic field of 300 Oe and at an incidence angle of 15°. The bottom of the figures presents the XMCD asymmetries.

### IV. X-ray absorption and magnetic circular dichroism of the Py/SLG sample

Figure S3(a) shows carbon K edge x-ray absorption spectra acquired with linear vertical polarization of the electric field for sample Py/SLG. Similarly to the results obtained for sample YIG/SLG (main text), the angular dependence of the XAS lineshape confirms that the $\pi^*$ unoccupied states (feature around 285 eV, C 1s→$\pi^*$ transition) are aligned out-of-plane. This result also corroborates the good quality of the SLG transferred to Py thin film.

Figure S3(b) shows XMCD asymmetry spectra measured at both Ni $L_{3,2}$ and carbon K edges. The Ni spectrum is representative of the Py layer and characteristic of the metallic environment of this material, whilst the C spectrum is somehow similar to that of the YIG/SLG, presenting a remarkable asymmetry around the feature related to the $\pi^*$ unoccupied states. This result is also in accordance with references[62] and[63] for Ni(111)/graphene, suggesting that the C $2p_z$ orbitals of the SLG are magnetically polarized due to the hybridization with the valence-band states of the adjacent Py layer.

### V. The EDX spectrum of graphene grown by CVD

The energy dispersive X-ray (EDX) spectrum of SLG by CVD onto the YIG film can be seen in Figure S4(a). The EDX spectrum taken from an arbitrary region of the sample shows the presence only of yttrium (Y), iron (Fe), oxygen (O) of the YIG film, and carbon (C) of SLG. Different samples were prepared, in order to confirm the results EDX measurements. In Figure S4(b), the samples for EDX measurements are prepared on a standard transmission electron microscope (TEM) nickel grids and are hence suspended graphene samples (as shown at the top of the insert). The image of the Ni-grid/SLG sample analyzed by electron microscopy is shown on the bottom of the insert of Figure S4(b). The EDX spectrum shows the presence only of oxygen (O) and carbon (C) of SLG. The additional peaks of the Ni, Al and Mg in the EDX spectrum are due to the presence of Ni TEM grids and the stub (compound of Al and Mg) that served as support on which the graphene samples are prepared for analysis. This result confirms that there are no visible traces of Cu adatoms on the graphene layer.

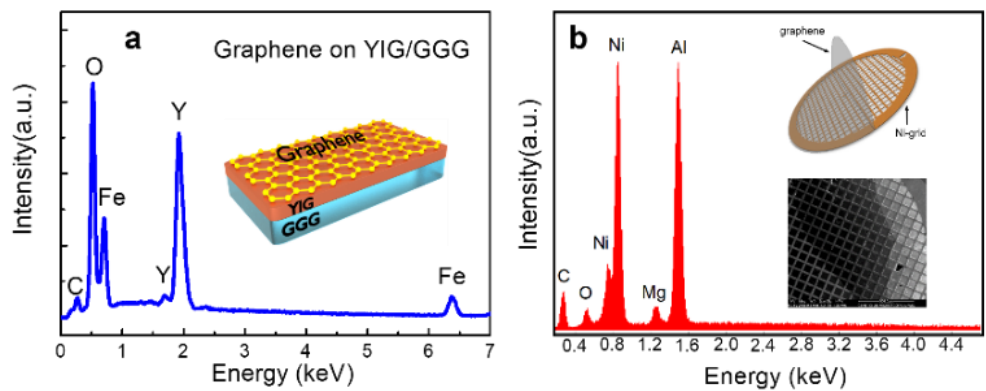

**Figure S4**. (color online) (**a**) EDX spectrum from an arbitrary region in the sample of GGG/YIG/graphene. (**b**) The EDX spectrum of SLG grown by CVD onto the nickel grids. The additional peaks of the Ni, Al and Mg in the EDX spectrum are due to the presence of Ni TEM grids and the stub (compound of Al and Mg).